\documentclass[letterpaper,11pt]{article}
\usepackage{pdfpages}
\usepackage{cite}
\usepackage[in]{fullpage}
\usepackage{graphicx,amssymb,amsmath,lineno}
\usepackage{setspace}
\usepackage{color}
\usepackage{xcolor}
\usepackage{upgreek} 

\title {\bf Harnessing an elastic flow instability to improve the kinetic performance of chromatographic columns}

\author {Fabrice Gritti$^{a,}$\thanks{Co-first author; Corresponding author: fabrice\textunderscore gritti@waters.com.},\ Emily Chen$^{b,}$\thanks{Co-first author.},\ Sujit S. Datta$^{b,}$\thanks{Corresponding author: ssdatta@princeton.edu.}  \\ $^a$Waters Corporation, Instrument/Core Research/Fundamental, \\ 34 Maple Street, Milford, MA, 01757, USA \\ $^b$Department of Chemical and Biological Engineering, \\ Princeton University,\\41 Olden Street, Princeton, NJ, 08544, USA}  

\date{}
\makeatletter

\def\@cite#1#2{$^{\mbox{\scriptsize #1\if@tempswa,#2\fi}}$}

\makeatother

\doublespace

\begin{document}

\maketitle\thispagestyle{empty}

\begin{abstract}
\noindent Despite decades of research and development, the optimal efficiency of slurry-packed HPLC columns is still hindered by inherent long-range flow heterogeneity from the wall to the central bulk region of these columns. Here, we show an example of how this issue can be addressed through the straightforward addition of a semidilute amount (500~ppm) of a large, flexible, synthetic polymer (18~MDa partially hydrolyzed polyacrylamide, HPAM) to the mobile phase (1\% NaCl aqueous solution, hereafter referred to as ``brine'') during operation of a 4.6 mm $\times$ 300 mm column packed with 10~$\mu$m BEH$^{\mathrm{TM}}$ 125~\AA \ Particles. Addition of the polymer imparts elasticity to the mobile phase, causing the flow in the interparticle pore space to become unstable above a threshold flow rate. We verify the development of this elastic flow instability using pressure drop measurements of the friction factor versus Reynolds number. In prior work, we showed that this flow instability is characterized by large spatiotemporal fluctuations in the pore-scale flow velocities that may promote analyte dispersion across the column. Axial dispersion measurements of the quasi non-retained tracer thiourea confirm this possibility: they unequivocally reveal that operating above the onset of the instability improves column efficiency by significantly reducing peak asymmetry. These experiments thereby provide a proof-of-concept demonstration that elastic flow instabilities can be harnessed to mitigate the negative impact of trans-column flow heterogeneities on the efficiency of slurry-packed HPLC columns. While this approach has its own inherent limitations and constraints, our work lays the groundwork for future targeted development of polymers that can impart elasticity when dissolved in commonly used liquid chromatography mobile phases, and can thereby generate elastic flow instabilities to help improve the resolution of HPLC columns.
\end{abstract} 
 
\noindent


\section{Introduction} 
Previous knowledge pertaining to the simulation of analyte dispersion along random sphere packings has demonstrated that the minimum reduced plate height (RPH) of an infinite diameter column—characterized by the absence of border and wall effects—should be around 0.9 when the particles are fully porous, 0.7 for superficially porous particles, and 0.5 for non-porous particles~\cite{Delgado, HlushkouUlrich, TallarekConfined1, KhirevichBulk, Daneyko2, Daneyko4, F109, F115, F149, FWaters21, FWaters23, FWaters24, KenModel}. From an experimental viewpoint, it is noteworthy that minimum RPHs close to 1.0 have already been observed, but under impractical conditions that cannot be implemented in production by manufacturers of modern HPLC and UHPLC columns. For instance, this was achieved by 1) slow and gentle dry-packing of very large 200-220 $\mu$m non-porous particles in a 2.4 cm inner diameter (i.d.) $\times$ 16.6 cm long column~\cite{Magnico}; 2) slurry-packing of capillary columns for a column-to-particle diameter ratio (or bed aspect ratio) smaller than 7 (12 $\mu$m i.d. capillary column packed with 5.0 $\mu$m porous particles~\cite{HsiehJorgenson}, 21 $\mu$m i.d. capillary column packed with 5.0 $\mu$m FPPs~\cite{KennedyJorgenson}, 3 mm i.d. column packed with 480 $\mu$m glass beads~\cite{KnoxParcher, Knox4}); 3) 4.6 mm i.d. $\times$ 10 cm long column packed with 2.5 $\mu$m superficially porous particles~\cite{F93}; and 4) 75 $\mu$m i.d. $\times$ 1 m long capillary column packed with 2 $\mu$m BEH Particles at very high pressures (30 000 psi), using highly concentrated slurry suspension (200 g/L) and by applying sonication (80 kHz)~\cite{Godinho}. In all of these cases, the bed structure is found to be quasi-uniform across the entire column diameter (no or little wall effects) but either the consolidated bed happens to be unstable, the column efficiency is too low (too large particle size), or the packing protocol is cumbersome and non-reproducible. To date, all modern HPLC columns massively adopted by the pharmaceutical industry are slurry-packed at very high pressure with 2 to 5 $\mu$m particles: their minimum RPHs are systematically reported in a narrow range from 1.7 to 2.1 for fully porous particles. 

\begin{figure}[tb]  
\begin{center} 
\caption{Schematic highlighting (a) the problem associated with radial flow heterogeneity in chromatography columns, and (b) our proposed solution using an elastic flow instability to promote analyte transverse dispersion. (a) Plot of the interstitial linear velocity $u(r)$, normalized by the center velocity $u_c$, within a 2.1~mm i.d. $\times$ 50~mm slurry-packed column with 2~$\mu$m particles~\cite{Arved}, as a function of the radial distance from the column wall $r$ normalized by the particle diameter ${d_p}$. The data are obtained using focused-ion-beam scanning electron microscopy and the solid curve is a best fit. The velocity drop-off near the wall is evidence of a wall-to-center velocity bias that negatively affects column performance. Adapted from~\cite{F124} with permission. (b) An elastic flow instability enhances transverse mixing of solute in a 3D porous medium: fluorescence confocal micrographs of solute (dyed) and solute-free (undyed) streams mixing. The polymer solution shows pronounced mixing due to an elastic instability (bottom row) compared to the case of a polymer-free solvent (top row). Images were taken at three different longitudinal distances along the porous medium equal to 0, 10, and 24 particle diameters. The average linear velocity is 1.7~mm/s. Reproduced with permission from~\cite{browne_harnessing_2023}.} \label{1}
\vskip 0.5in 
\begin{minipage}[t]{0.99\linewidth}  
\includegraphics[width=\linewidth]{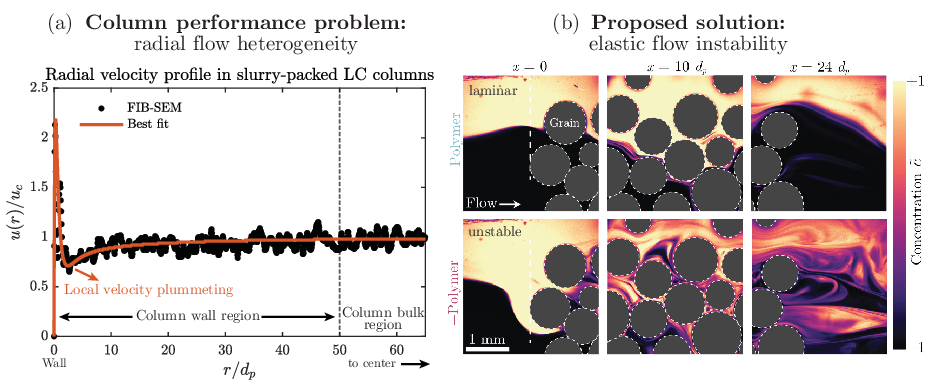}
\end{minipage}
\end{center} 
\end{figure}

Such a significant performance gap between the idealized and practical RPH (from 0.9 to 1.7-2.1 for porous particles) has been unequivocally explained from the flow unevenness of the mobile phase across the column i.d. or the so-called long-range eddy dispersion in chromatography~\cite{Kaminski, Farkas, Roper, Frey, Baur, Abia, F90, Astrath, F85, F82, HsiehJorgenson, Arved, Bruns2, BrunsSegregation, AndrewRev} [Figure~\ref{1}(a)]. The transverse flow heterogeneity in packed columns is directly related to the variation of the local bed density which itself results from the radial distribution of shear stress applied to the particles during slurry-packing and bed consolidation~\cite{YewGuiochon1, YewGuiochon2, Shalliker2, Shalliker3}. The stress experienced by the particles in the wall region happens to be higher than that in the bulk central region of the column because of the presence of the fixed column wall~\cite{YewGuiochon2}. This effect was directly observed by focused ion beam scanning electron microscopy for 2 $\mu$m BEH-C$_{18}$ Particles slurry-packed in 2.1 mm i.d. $\times$ 50 mm stainless steel column~\cite{Arved} and indirectly revealed by local electrochemical detection across the outlet column diameter~\cite{Jude1, Jude2, F90, F97}.

To date, research efforts to bridge this performance gap have only concerned small scale (nanoflow) empty channels and open tubular capillary columns by enhancing the transverse transport of the analyte over the surface perpendicular to main flow direction. Indeed, the RPH associated with long-range eddy dispersion is inversely proportional to the transverse dispersion coefficient~\cite{Golay, F115, FWaters24}, where transverse dispersion is only generated by natural diffusion and transverse convection induced by external stationary phase obstructions in packed columns~\cite{Daneyko2}. Solutions based on the application of turbulent flow at high Reynolds numbers~\cite{GiddingsScience, Pretorius, Stovin, Cramers, Knox, MartinGuiochon1, FWaters14, FWaters16}, alternating current electro-osmetic flow (AC-EOF)~\cite{MartinGuiochon2, MartinGuiochon3, DeMalshe2,Bihi}, and acoustic streaming~\cite{DeMalshe1} have been explored in separation systems. However, they cannot be scaled to large-scale systems such as 4.6 mm i.d. HPLC columns packed with sub-10 $\mu$m particles and they may require active electrical supply in the case of acoustics and AC-EOF. Therefore, a complete passive solution applicable to large-volume chromatographic systems is still needed.

One potential passive solution to enhance transverse dispersion of analytes is the use of an elastic flow instability~\cite{larson_instabilities_1992,shaqfeh_purely_1996, pakdel_elastic_1996,groisman_elastic_2004,browne_bistability_2020,browne_elastic_2021,DattaReview,browne_homogenizing_2023,browne_harnessing_2023} generated by the addition of flexible high molecular weight polymers to the mobile phase, which imparts elasticity to the fluid even in dilute amounts. As the polymer chains traverse tortuous, curved streamlines within the porous medium, elastic stresses can accumulate and destabilize the flow, generating turbulent-like chaotic fluctuations despite negligible inertia (Reynolds number, $\mathrm{Re} \ll 1$). Recent work by Browne \& Datta~\cite{browne_elastic_2021} directly observed and characterized the onset of such an elastic instability during the flow of a dilute solution of partially hydrolyzed polyacrylamide (HPAM) in a three-dimensional (3D) packing of consolidated glass spheres (average diameter $\approx330~\mu$m). The instability is characterized by persistent spatiotemporal flow fluctuations above a threshold flow rate or Weissenberg number, $\textrm{Wi}$, which describes the ratio of elastic to viscous stresses in the fluid. Further, the onset of the instability coincides with a large increase in the flow resistance across the medium, resulting from an increase in energy dissipation from the elastic instability. Thus, one indicator of an elastic instability—without having access to the pore-scale flow fields—is an increase in the macroscopic flow resistance.

Flow fluctuations associated with the onset of an elastic instability have tantalizing potential for use as a mechanism to enhance mass, solute, and heat transport in low Reynolds number porous media flows where transport is typically dominated by slow diffusion. Indeed, previous work has shown that elastic instabilities may enhance solute mixing and increase the rate of chemical reactions~\cite{groisman_efficient_2001,burghelea_chaotic_2004,browne_harnessing_2023}, redirect flow to lower permeability regions in stratified porous media~\cite{browne_homogenizing_2023}, alter transverse and axial dispersion properties of colloids~\cite{scholz_enhanced_2014,Guasto2023}, and improve heat transfer efficiency~\cite{traore_efficient_2015,whalley_enhancing_2015}. An example is shown in Figure~\ref{1}(b), which presents confocal fluorescence images taken within a porous medium, showing enhanced solute mixing and dispersion generated by an elastic instability. However, to the best of our knowledge, elastic flow instabilities have yet to be explored in large-scale porous systems such as slurry-packed chromatographic columns. Thus, research is needed to investigate whether elastic instabilities can help improve the resolving power of chromatography columns. 
 
In this work, we demonstrate the first, to our knowledge, proof-of-concept application of an elastic flow instability in the interparticle volume of slurry-packed HPLC columns in order to increase the column's resolving power. The method consists of adding HPAM to the mobile phase to generate the elastic instability. First, we measure the bulk rheological properties of a 500 ppm solution of 18 MDa HPAM dissolved in brine to determine the bulk elastic properties of the mobile phase and thereby predict the flow conditions necessary for the elastic instability (Wi $>$ Wi$_c$), where Wi$_c$ represents a threshold value of the Weissenberg number. Secondly, we verify that such elastic instabilities arise in a 4.6 mm $\times$ 300 mm column packed with 11.3 $\mu$m BEH 125 \AA \ Particles using friction factor versus Reynolds number measurements. Finally, we perform axial dispersion measurements of a thiourea analyte, which demonstrate that the concurrent injection of polymer in the mobile phase does indeed mitigate the negative impact of trans-column flow heterogeneities on the efficiency of the HPLC columns when operating above the onset of elastic instability. Our experiments therefore highlight the potential of elastic instabilities for improving column performance; however, the important constraints of detection compatibility, chemical inertness, and chromatographic speed required by modern routine HPLC applications should be considered in future research.

\section{Theoretical details}  \label{Theory}
\subsection{Mobile phase rheology} \label{theory:rheology}
We use the Carreau-Yasuda constitutive model to describe the steady-shear rheology of the polymer solution~\cite{larson_structure_1999}. In particular, we fit our measurements of the dynamic shear viscosity using the relation:
\begin{equation}
\eta(\dot{\gamma})=\eta_{\infty}+(\eta_0-\eta_{\infty})\left(1+\left(\frac{\dot{\gamma}}{\dot{\gamma}_c}\right)^a\right)^{\frac{n-1}{a}},
    \label{eq:CY}
\end{equation}
where $\dot{\gamma}$ is the imposed shear rate, $\eta_0$ is the zero shear viscosity, $\eta_{\infty}$ is the infinite shear viscosity, $\dot{\gamma}_c$ is the onset shear rate for shear-thinning, $a$ is a parameter ($a \simeq 2$) that controls the transition from the low shear rate Newtonian plateau to the shear thinning regime, and $n$ is the power law index ($0\leq n \leq 1$). At low shear rates, the Carreau-Yasuda fluid behaves like a Newtonian fluid with constant viscosity $\eta_0$; as the shear rate increases beyond $\dot{\gamma}_c$, the fluid begins to shear-thin with a power law decay described by $n$. At high shear rates, the Carreau-Yasuda fluid again behaves like a Newtonian fluid with viscosity $\eta_{\infty}$. Physically, the shear-thinning of the polymer solution is due to microstructural rearrangement and alignment of the polymer chains under flow, leading to a reduction in the solution viscosity~\cite{macosko_rheology_1994,larson_structure_1999,rubinstein_polymer_2003,ryder_shear_2006}.

In addition to the shear viscosity, we simultaneously measure the first normal stress difference, $N_1(\dot{\gamma}) = \tau_{xx}-\tau_{yy}$, where $\tau_{xx}$ is the streamwise normal stress in the direction of flow and $\tau_{yy}$ is the transverse normal stress (i.e. radial direction in a cylindrical geometry). We fit the experimental $N_1$ flow curves to a power law, $N_1(\dot{\gamma})=K_{N_1}\dot{\gamma}^{n_{N_1}}$, 
where $K_{N_1}$ is the material prefactor with units $\mathrm{Pa\cdot s^{n_{N_1}}}$ and $n_{N_1}$ is the power law index.

\subsection{Flow through porous HPLC columns}
The flow is described by two key dimensionless parameters: the Reynolds number quantifying the relative strength of inertial to viscous stresses, and the Weissenberg number quantifying the relative strength of elastic to viscous stresses. The former is defined as $\mathrm{Re}=\frac{\rho u_S d_p}{\eta (1-\epsilon_e)}$,
where $\rho$ is the density of the fluid, $u_S$ the average superficial linear velocity (flow rate per unit area of the empty column), $d_p$ the average particle size, $\eta$ the fluid dynamic shear viscosity, and $\epsilon_e\simeq$ 40\% the external porosity of the column packing. The latter is defined as 
$\mathrm{Wi}=\tau \dot{\gamma}$, where $\tau$ is the longest relaxation time of the polymer solution, $\dot{\gamma}= \frac{16}{3}\frac{u_S}{\epsilon_e d_{\mathrm{tube}}}$ is the average shear rate in the pore space~\cite{FWaters67}, and\, $d_{\mathrm{tube}}=0.42\frac{\epsilon_e}{1-\epsilon_e}d_p$ is the equivalent hydraulic diameter in a chromatographic column~\cite{Wan}.

The pressure drop per unit of column length can be described using the standard Ergun equation~\cite{Ergun}, modified to also include energy dissipated by the elastic instability following our previous work~\cite{browne_elastic_2021, browne_homogenizing_2023}:

\begin{equation} \label{Ergun}
\frac{\Delta P}{L}=K_c\frac{(1-\epsilon_e)^2}{\epsilon_e^3d_p^2} \eta(\dot{\gamma}) u_S + 1.75 \frac{1-\epsilon_e}{\epsilon_e^3d_p} \rho u_S^2 + \frac{\Delta P_{EI}}{L},   
\end{equation}
where $\Delta P$ is the pressure drop along the column, $L$ its length, $K_c$ an empirical constant ($\simeq$ 180 for the specific tortuosity and surface area of consolidated random sphere packings), and $\Delta P_{EI}\geq0$ the added pressure drop due to the elastic instability.
The first and second terms on the right-hand-side of Eq.~\ref{Ergun} account for the resistance to fluid flow due to dissipation by viscous and inertial forces, respectively, in the stable laminar base case; for the range of Reynolds numbers investigated in this work ($\mathrm{Re} \ll 1$), the second term in Eq.~\ref{Ergun} is always negligible, and we therefore do not consider it hereafter. The third term on the right-hand-side of Eq.~\ref{Ergun} generally accounts for the added resistance imparted by the elastic flow instability; it can be calculated by considering the viscous dissipation associated with spatio-temporal fluctuations of the strain-rate tensor~\cite{browne_elastic_2021}, and increases with increasing $u_S$ when Wi $>$ Wi$_c$.

Following convention, we then define the friction factor as:
\begin{equation} \label{f}
f=\frac{\Delta P}{L}\frac{d_p}{\rho u_S^2}\frac{\epsilon_e^3}{1-\epsilon_e}.
\end{equation}
In the case of laminar flow ($\mathrm{Re}\ll1$ and $\mathrm{Wi} \ll 1$)  with a Newtonian fluid ($\eta$ is shear rate independent), the combination of Eq.~\ref{f} and simplified Eq.~\ref{Ergun} gives
\begin{equation} \label{fNewtonian}
\log(f)= \log(K_c) - \log(Re).
\end{equation} 
Therefore, in a log-log representation, the friction factor is expected to decrease linearly with increasing Reynolds number; any deviation from this linearity reveals a change in flow resistance relative to Newtonian fluids under laminar flow. This deviation can be due to e.g., a modification of the flow path tortuosity, a change in external porosity $\epsilon_e$, turbulent flow at $\mathrm{Re} > 2000$, or an elastic instability at Wi $>$ Wi$_c$.

\subsection{Resolution of HPLC columns}  
The resolution of a chromatographic column can be adversely impacted by the existence of long-range velocity biases from the bulk center region to the wall region [Figure~\ref{1}(a)]. However, the transverse dispersion of analyte across the column diameter~\cite{Daneyko2} mitigates the negative impact of such radial flow heterogeneities on the column efficiency because it permits the analyte to be exchanged between different flow streamlines. 

We quantify this intuition using a simple calculation, defining the absolute radial position, $r$; column inner radius, $r_c$; dimensionless radial position across the column, $x=\frac{r}{r_c}$; column plate height associated with long-range eddy dispersion, $H_{\mathrm{eddy, long}}$; effective diffusion coefficient of the analyte in the packed bed volume in the absence of convection, $D_{\mathrm{eff}}$; local migration velocity of the analyte at radial position $x$, $u_R(x)$; radially-averaged migration linear velocity of the analyte, $\bar{u}_R$; and the local transverse dispersion coefficient of the analyte at radial position $x$, $D_t(x)$. When the number of analyte exchanges becomes very large (infinitely long column, asymptotic dispersion regime), the column plate height is given by extending~\cite{ArisGeneral} the general dispersion theory of Aris for open tubular columns to packed beds by considering their specific radial velocity profile~\cite{F69, F115, Arved} [Figure~\ref{1}(a)]. In particular, 
\begin{equation} \label{HeddyLong}
H_{\mathrm{eddy, long}}= A_0 \frac{4r_c^2}{D_{\mathrm{eff}}} \bar{u}_R,
\end{equation}   
where the coefficient~\cite{F115} $A_0=\frac{I_1 - 2 I_2 + I_3}{2}$ with
\begin{eqnarray}
I_1 & = & \int_0^1 \frac{\Phi^2(x)}{2x\Psi(x)} \mathrm{d}x \label{I1} \\
I_2 & = & \int_0^1 \frac{x\Phi(x)}{2\Psi(x)} \mathrm{d}x \label{I2} \\
I_3 & = & \int_0^1 \frac{x^3}{2\Psi(x)} \mathrm{d}x  \label{I3},
\end{eqnarray} 
$\Phi(x)=\int_0^x 2x^{'}\phi(x^{'})\mathrm{d}x^{'}$, $\phi(x)=\frac{u_R(x)}{\bar{u}_R}$, and $\Psi(x)=\frac{D_t(x)}{D_{\mathrm{eff}}}$. Hence, the plate height $H_{\mathrm{eddy, long}}$ can be reduced by promoting transverse dispersion, i.e., increasing $D_t(x)$, which appears in the denominators of the right-hand-side integrals in Equations~\ref{I1},~\ref{I2}, and~\ref{I3}, thereby reducing $A_0$. Promoting transverse dispersion of the analyte is therefore a useful way to limit the negative impact of radial flow heterogeneity on the performance of chromatographic separation systems~\cite{F63, ParmentierCR, FWaters51, FWaters52}. This is the main rationale of our work, where our objective is to determine whether an elastic flow instability induced by the addition of a well-chosen polymer in the mobile phase can help mitigate the negative impacts of radial flow heterogeneities on column efficiency by promoting transverse dispersion.

\section{Experimental details}

\subsection{Chemicals}
We use HPLC grade water from Fisher Scientific (Fair Lawn, NJ, USA) as the solvent; the added sodium chloride salt (NaCl) and thiourea ($>$99\% purity) are obtained from Sigma Aldrich (Milwaukee, WI, USA). The polymer additive used is 18 MDa partially hydrolyzed polyacrylamide (HPAM, 30\% hydrolysis) obtained from Polysciences, Inc (Warrington, PA, USA).

\subsection{Instrumentation} \label{1290}
We acquire all chromatographic data on the Arc System$^{\mathrm{TM}}$ (Waters, Milford, MA, USA). The standard configuration of this instrument includes a multi-solvent delivery system (the quaternary solvent manager, QSM), a gradient proportioning valve (GPV), a primary and accumulator pump heads, a two-paths flow mixer, an auto-sampler with a 30~$\mu$L sample loop, a two-column selection valve, a two-column semi-adiabatic oven, and a single wavelength transmission UV-detector (8~$\mu$L cell volume). In this work, we place the QSM pump with an isocratic solvent manager (ISM) pump equipped with primary and accumulator pump heads, but without any mixer.  

The instrument is run by the Empower$^{\mathrm{TM}}$ 3 Chromatography Data Software (Waters, Milford, MA, USA). The extra-column volumes are 0.042 mL from the auto-sampler needle seat to the UV cell. The offset time between the moment the chromatogram starts being recorded and the moment the injection valve actually switches is 0.7~s. All measurements are carried out at room temperature (297$\pm$1~K). The wavelength of the UV detector is set at 272~nm, with a sampling rate of 10~Hz.

\subsection{Columns} \label{Columns} 
The research prototype column (4.6~mm $\times$ 300~mm) is home-packed with 11.3~$\mu$m  fully porous BEH125 Particles prepared on-site (Waters, Milford, MA). We determine the characteristic particle diameter, 11.3~$\mu$m, by identifying the median of the particle size distribution measured using a Coulter counter. The external porosity of the column is 42.3\%, as determined using permeability measurements (pure water, $T=297$~K, $\eta$=0.91~mPa$\cdot$s, $K_c$=180) performed at flow rates of 0.1, 0.25, 0.50, 0.75, and 1.50~mL/min.

\subsection{Bulk rheology of the HPAM-modified mobile phase} 
The polymer-amended mobile phase is a 500~ppm aqueous solution of the 18 MDa HPAM polymer with 1~wt\% NaCl added. To mix the polymer solution, we gradually add the HPAM in stepwise increments to water while stirring gently with a magnetic stir bar for 24~h. We then add the NaCl salt in stepwise increments and allow the solution to mix for another 24~h. We then let the solution rest undisturbed to allow for bubbles to dissipate. All polymer solutions are used within 1 month of preparation. The overlap concentration describing the concentration at which adjacent HPAM polymer coils begin to interact in solution is $c^* = 200$~ppm~\cite{liu_concentration_2009}. Our experimentally-chosen concentration of 500 ppm thus corresponds to the semi-dilute regime, with $c/c^* = 2.5$. 

We characterize the rheology of the polymer-amended mobile phase using a stress-controlled Anton Paar MCR501 rheometer (Ashland, VA, USA) fitted with a truncated cone-plate geometry (CP50-2: 50~mm diameter, 2$^{\circ}$, 53~$\upmu$m gap), operating at 298~K. The measurements are presented and detailed in \S\ref{Rheology}. We measure steady-state flow curves for the polymer solution by ramping down over shear rates of $\dot{\gamma}=100-0.1 \: \mathrm{s^{-1}}$, with the time-per-shear rate set by the instrument. We measure four replicates of separate samples from the same polymer solution batch and report the standard deviation as error bars.

\subsection{Column friction factor and efficiency measurements} \label{friction} 
We characterize the performance of our prototype column by measuring the friction factor $f$ and efficiency $N$ at a temperature $T=297$~K at various stages of operation. To determine $f$, we perform pressure drop measurements over a range of flow rates and apply Eq.~\ref{f}. To determine $N$, we measure the first ($\mu_1$) and second ($\mu_2^{'}$) central moments of the concentration distribution recorded at a detection wavelength of 272~nm (10 Hz sampling rate) for a 5~$\mu$L volume of the small molecule thiourea injected at a flow rate of $F_v=0.5$~mL/min. When the thiourea is dissolved in brine, this flow rate corresponds to conditions of laminar steady flow in the column; however, when the thiourea is dissolved in HPAM-amended brine (step \#3 below), this flow rate is well above the threshold of elastic instability, as described further in \S\ref{Results}. To account for retention and dispersion taking place in the HPLC system (injection needle, needle seat capillary, inlet and outlet connecting tubes, and detection cell volumes), we correct the moments $\mu_1$ and $\mu_2^{'}$ by those measured in the same conditions but after replacing the column with a zero dead volume union connector, $\mu_{1,ex}$ and $\mu_{2,ex}^{'}$; the efficiency is thus given by $N=\frac{\left(\mu_1-\mu_{1,ex}\right)^2}{\mu_2^{'}-\mu_{2,ex}^{'}}$. 

We thereby characterize the column performance through the following series of experiments:
\begin{enumerate}
\item Equilibrate the column with a mobile phase of just brine for 1.5~h at a flow rate of 1~mL/min, corresponding to $\simeq$18 empty column volumes. We then increase the flow rate stepwise (0.1 mL/min step, each plateau lasting 2~min) from 0.1 to 2.0~mL/min. This measurement thereby provides the reference flow resistance of the column for a Newtonian fluid. We also use the thiourea concentration measurements to determine the baseline efficiency of the pristine column before contact with the HPAM polymer, $N=N_{\mathrm{Ref}}$.

\item Equilibrate the column with the HPAM-amended mobile phase  for 15~h at a flow rate of 0.1~mL/min, again corresponding to $\simeq$ 18 empty column volumes. We then wash the column abundantly with the polymer-free brine and repeat the thiourea concentration measurements in brine to determine how the column efficiency is altered by contact with the HPAM polymer, $N=N_{\mathrm{Before}}$.

\item Again equilibrate the column with the HPAM-amended mobile phase  for 15~h at a flow rate of 0.1~mL/min, again corresponding to $\simeq$ 18 empty column volumes, and then increase the flow rate stepwise from 0.0125 to 0.025, 0.05, 0.075, 0.100, 0.125, 0.150, 0.175, 0.200, 0.225, 0.250, and 0.275~mL/min. For each 60~min-long flow step, the flow rate increases linearly with time during the first 5~min and is then maintained constant over the remaining 55~min in order to reach a steady pressure drop along the column. This measurement thereby provides the flow resistance of the column for the polymer-amended fluid. We also use the thiourea concentration measurements, this time in the HPAM-amended mobile phase, to determine the column efficiency in the elastically-unstable flow state, $N=N_{\mathrm{HPAM}}$.

\item Wash and re-equilibrate the column with the polymer-free brine for 3~h at a flow rate of $F_v=0.5$~ml/min, again corresponding to $\simeq$18 empty column volumes, until the pressure drop returns to a new steady state pressure. The flow rate is then increased stepwise (each plateau lasting 2~min) from 0.0125 to 0.025, 0.05, 0.075, 0.100, 0.125, 0.150, 0.175, 0.200, 0.225, 0.250, and 0.275~mL/min. This measurement provides the flow resistance of the column after using the polymer-amended fluid. We repeat the thiourea concentration measurements in polymer-free brine to determine the column efficiency after using the polymer-amended fluid, $N=N_{\mathrm{After}}$. Together, these measurements characterize whether the column is modified in any way by the polymer.
\end{enumerate}

\section{Results} 
\label{Results}
\subsection{Rheology}  \label{Rheology}
\begin{figure}[tb]  
\begin{center} 
\caption{Polymer solution rheology. (a) Flow curve of dynamic viscosity, $\eta(\dot{\gamma})$, for the 500 ppm HPAM in brine solution, exhibiting strong shear-thinning behavior. Empty circles correspond to averages over 4 separate measurements and error bars reflect one standard deviation. The solid curve is the best fit to the Carreau-Yasuda model. We show an extrapolation of the Carreau-Yasuda model to the relevant shear rate regime encountered in HPLC applications, suggesting that the polymer solution viscosity lies in the high shear-rate plateau. (b) First normal stress difference, $N_1(\dot{\gamma})$, versus applied shear rate. The solid curve is the best fit to a power law model with power law exponent of $\alpha_{N_1} = 0.85$.} \label{2}
\vskip 0.5in 
\begin{minipage}[t]{0.99\linewidth}  
\includegraphics[width=\linewidth]{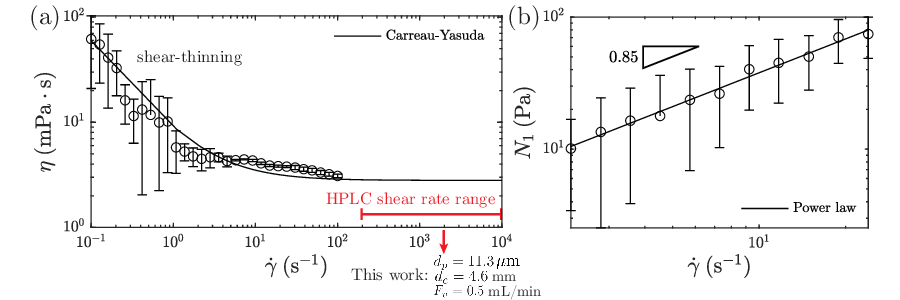}
\end{minipage}
\end{center} 
\end{figure}

The results of the rheology measurements of the HPAM polymer-amended mobile phase are shown in Figure~\ref{2}. Figure~\ref{2}(a) shows the steady-shear viscosity flow curve, $\eta(\dot{\gamma})$. The polymer solution is already strongly shear-thinning at the lowest measurable shear rates in the rheometer. The solid curve represents the best fit of the standard Carreau-Yasuda model (Eq.~\ref{eq:CY}) with $\eta_0 = 1.47 \: \mathrm{Pa\cdot s}$, $\eta_{\infty}=2.8 \: \mathrm{mPa\cdot s}$, $\dot{\gamma}_c = 0.0033 \: \mathrm{s^{-1}}$, and $n = 0.05$, with $a$ set to 2. We extrapolate this fit to larger shear rates $
\sim10^4 \: \mathrm{s^{-1}}$ to compare to shear rates typically encountered in HPLC applications; as shown by the red line in Figure~\ref{2}(a), the corresponding polymer solution viscosity expected for our solution under typical HPLC operating conditions lies in the high shear rate Newtonian plateau, $\eta_{\infty}$. In particular, in the experiments using thiourea concentration measurements to determine column efficiency, we use an imposed constant flow rate of $F_v = 0.5$~mL/min, which corresponds to a shear rate of $\dot{\gamma} \approx 2000 \: \mathrm{s^{-1}}$ and $\mathrm{Wi}=360$. This Wi is well above the threshold of elastic instability, Wi$_c \approx 85$, determined from the onset of flow-thickening in the friction factor measurements in Figure~\ref{3}(b).

Figure~\ref{2}(b) shows the first normal stress difference, $N_1(\dot{\gamma})$, and corresponding power law fit, $N_1(\dot{\gamma}) = K_{N_1}\dot{\gamma}^{n_{N_1}}$ with $K_{N_1} = 5.3 \; \mathrm{Pa \cdot s}^{n_{N_1}}$ and $n_{N_1} = 0.85$. The lower limit of the measured first normal stress difference, $N_1$, is approximately 10~Pa due to the normal force sensitivity of the instrument, limiting the range of reported shear rate values for $N_1$. The growth of $N_1$ with increasing shear rate results from the development of elastic stresses as polymers undergo shearing-driven alignment in the curved geometry.

We could not obtain shear relaxation time measurements for the polymer-amended mobile phase because of instrument sensitivity limitations for HPAM concentrations below 1000~ppm. Therefore, to estimate the characteristic relaxation time of the polymer solution, we first calculate the Rouse and Zimm time scales for dilute polymer solutions. The Rouse time scale, which neglects intramolecular hydrodynamic interactions, is given by~\cite{rouse_theory_1953,zimm_dynamics_1956,chauveteau_thickening_1984,rubinstein_polymer_2003} $\tau_R = \frac{6}{\pi^2} \frac{\eta_s[\eta]M}{RT}=12.4$~ms, where $\eta_s$ is the solvent viscosity, $[\eta]$ is the intrinsic viscosity, $M$ is the polymer molecular weight, $R$ is the ideal gas constant, and $T$ is temperature. The Zimm time scale, which incorporates intramolecular hydrodynamic interactions, is given by~\cite{liu_concentration_2009} $\tau_Z = 0.422 \frac{\eta_s[\eta]M}{RT}=8.6$~ms. However, these time scales describe dilute polymer solutions, whereas our polymer-amended mobile phase is in the semidilute regime; the Rouse and Zimm time scales therefore represent lower bounds to the true solution relaxation time. Moreover, previous shear stress relaxation measurements~\cite{liu_concentration_2009} on a more concentrated solution of 1000~ppm HPAM in brine provide an upper-bound estimate of $\approx200$~ms for the true relaxation time. Hence, we expect that the solution relaxation time is somewhere in the range $\sim10-200$~ms; for simplicity, we use $\tau=200$~ms in calculating the Weissenberg number, but note that the true value of Wi may be lower by as much as a factor of 20.

\subsection{Column flow resistance} \label{Flow}
\begin{figure}[tb]  
\begin{center} 
\caption{Friction factor $f$ before, during, and after flow of the polymer-amended mobile phase determined from measurements of the pressure drop across the column using Eq.~\ref{f}. (a) Polymer-free 1\% NaCl aqueous mobile phase (brine) before the column is in contact with the HPAM polymer. The black curve shows the prediction for a Newtonian fluid in laminar flow ($f \sim \mathrm{Re}^{-1}$, Eq.~\ref{fNewtonian}) with $K_c$=180. (b) Polymer-amended (500~ppm of 18~MDa HPAM) brine mobile phase. The black curve shows the prediction for a Newtonian fluid in laminar flow with $K_c$=420. The slight vertical shift from the polymer-free case (indicated by the lower double-headed arrow) reflects irreversible adsorption of the polymer on the solid particles. The pronounced positive deviation from the red curve (indicated by the upper double-headed arrow) shows evidence of added flow resistance due to an elastic flow instability. (c) Polymer-free brine mobile phase after the column is in contact with the HPAM polymer, again showing Newtonian laminar behavior, confirming that the positive deviation from the red curve in (b) is due to the elastic flow instability and that the slight vertical shift from the black curve reflects irreversible polymer adsorption.} \label{3}
\vskip 0.5in 
\begin{minipage}[t]{0.99\linewidth}  
\includegraphics[width=\linewidth]{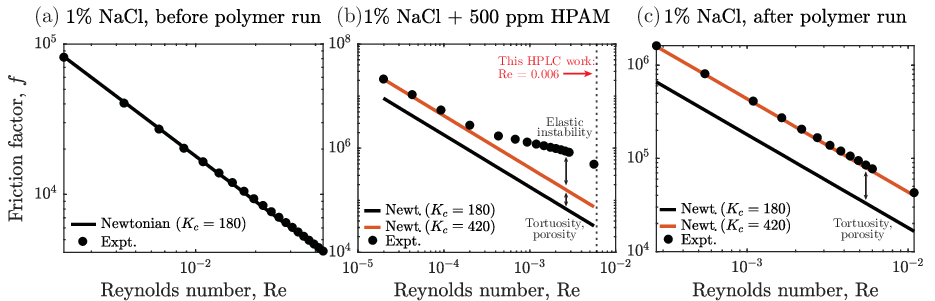}
\end{minipage}
\end{center} 
\end{figure}

Figure~\ref{3} summarizes our measurements of the flow resistance, quantified by the friction factor $f$, as a function of Reynolds number Re in the lab-scale test column for the different mobile phases used in the experiments: polymer-free brine and 500~ppm HPAM-amended brine. Figure~\ref{3}(a) shows the reference flow resistance before the packed BEH125 Particles are exposed to the HPAM polymer solution (step \#1 in \S\ref{friction}). As expected, the polymer-free brine is a Newtonian fluid of constant viscosity and the flow is in the laminar regime (0.002 $< \mathrm{Re} <$ 0.044): the experimental (circles) and theoretical (Eq.~\ref{fNewtonian} with $K_c=180$, solid line) plots of $\log f$ versus log Re are nearly indistinguishable.

We observe two key differences when using the HPAM-amended mobile phase (step \#3 in \S\ref{friction}), shown by the symbols in Figure~\ref{3}(b). First, the friction factor is shifted to higher values at the lowest Re, with the permeability constant $K_c$ increasing from 180 to 420 [red curve]. One possible explanation for this shift is that the HPAM adsorbs as a $\approx250~\upmu$m-thick film uniformly over the entire surface of the BEH125 Particles, reducing the overall packing porosity $\epsilon_e$ from $\approx42\%$ to $\approx35\%$. If true, the internal volume of the BEH Particles would become inaccessible and the elution times of the non-retained peaks of thiourea and chloride anions would be as small as 3.45~min; however, we consistently measure their mean elution time as 7.84~min, as detailed below in \S\ref{Knetics}. The measurement thus suggests that HPAM adsorption does not completely coat the BEH125 Particles and render them impermeable, but instead simply constricts the narrowest interparticle flow paths, augmenting the tortuosity of the flow-through channels by 53\%. This result is consistent with the findings of more simplified microfluidic experiments~\cite{parsa2020origin}. Second, despite the shear-thinning behavior of the bulk mobile phase [Figure~\ref{2}(a)], we observe flow-thickening at sufficiently large flow rates: As Re increases from 0.0001 to 0.0049 (Wi increases from $\approx$9 to 200), the measured friction factor exceeds the expected Newtonian behavior, indicated by the upper double-headed arrow in Figure~\ref{3}(b). This enhanced flow resistance does not arise due to inertial effects, given that $\mathrm{Re} \ll 1$. Instead, it reflects the onset of an elastic instability induced by the addition of HPAM~\cite{browne_elastic_2021}. We thus determine the critical Weissenberg number for the onset of the elastic instability from the onset of flow-thickening in the friction factor as $\mathrm{Wi}_\mathrm{c} = 85$. 

Finally, repeating the flow resistance experiments after abundantly washing the column with brine solution (step \#4 in \S\ref{friction}) provides a way to assess irreversible changes to the column caused by interactions with HPAM. As shown in Figure~\ref{3}(c), the flow resistance of the Newtonian brine solution remains shifted vertically relative to that measured prior to HPAM introduction, with $K_c$ remaining equal to 420 [red curve]---suggesting that HPAM adsorption to the external BEH surfaces is irreversible. This suggestion is corroborated by the thiourea elution experiments described below in \S\ref{Knetics}. Nevertheless, as expected, the plot shows Newtonian behavior in the absence of HPAM in the mobile phase.

\subsection{Mass transfer within the column} \label{Knetics} 
\begin{figure}[tb]  
\begin{center} 
\caption{Chromatograms of thiourea before (black) and after (blue) the column is equilibrated with the polymer-amended brine mobile phase and then washed with the polymer-free brine. In both cases, thiourea exhibits the same overall retention time, yet has a broader peak after the column has been equilibrated with the polymer, indicating irreversible but not complete adsorption of the polymer on the particle surfaces.} \label{4}
\vskip 0.5in 
\begin{minipage}[t]{0.99\linewidth}  
\includegraphics[width=\linewidth]{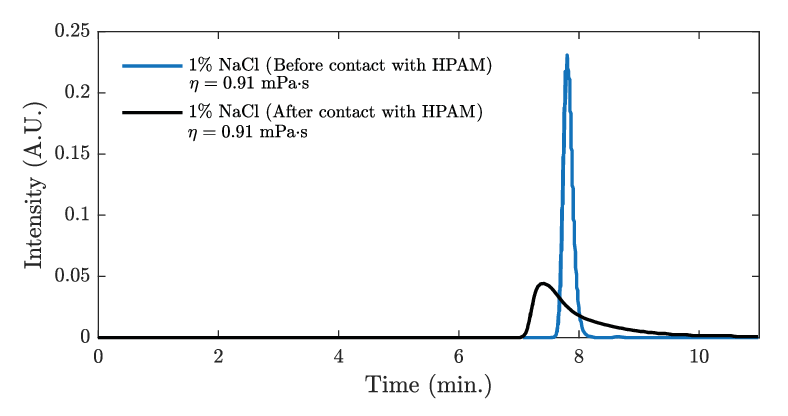}
\end{minipage}
\end{center} 
\end{figure}

\begin{figure}[tb]  
\begin{center} 
\caption{Normalized chromatograms of thiourea before (black), during (red), and after (blue) addition of HPAM polymer to the mobile phase. The elution curves are normalized by the peak intensity to facilitate comparison of their shapes. When the experiment is performed with polymer-amended mobile phase at a Weissenberg number (Wi) sufficiently large to generate an elastic instability, the elution peak is considerably more symmetric (red)---demonstrating that the elastic instability enhances transverse dispersion and increases column efficiency. } \label{5}
\vskip 0.5in 
\begin{minipage}[t]{0.99\linewidth}  
\includegraphics[width=\linewidth]{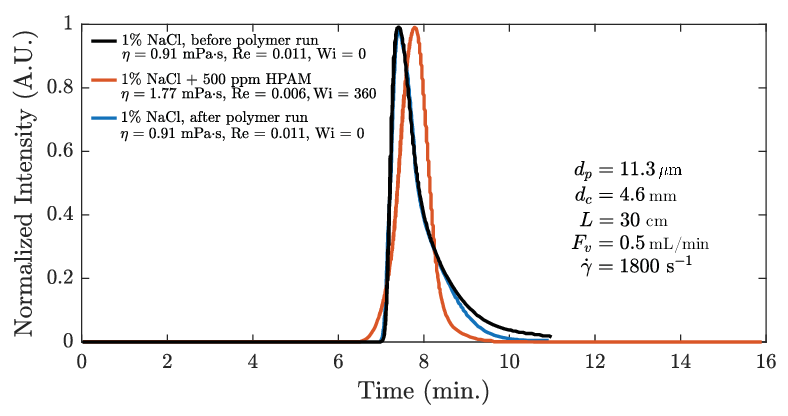}
\end{minipage}
\end{center} 
\end{figure}

We evaluate the column efficiency by measuring chromatograms of the non-retained compound thiourea. Before the packed BEH125 Particles are exposed to the HPAM polymer solution (step \#1 in \S\ref{friction}), the thiourea chromatogram shows a well-defined, approximately symmetric peak with an average retention time of 7.84~min [blue curve in Figure~\ref{4}, $N_{\mathrm{Ref}}$=8050]. After the particles are incubated with the HPAM-amended mobile phase (step \#2 in \S\ref{friction}), the chromatogram---obtained again with thiourea dissolved in the polymer-free brine mobile phase---shows the same average retention time of 7.84~min, but with a broader peak and lower efficiency [black curve in Figure~\ref{4}, $N_{\mathrm{Before}}$=160]. The same curve, normalized by its peak intensity, is shown by the black curve in Figure~\ref{5}. Notably, the unchanged average retention time demonstrates that the thiourea can still penetrate the same internal mesoporous volume of the BEH125 Particles; that is, the HPAM adsorption indicated by the flow resistance measurements described in \S\ref{Flow} does not completely coat the BEH125 Particles and render them impermeable. Instead, the broader peak shape and lower column efficiency suggests that irreversibly-adsorbed HPAM occludes some of the mesopore openings between particles, augmenting the tortuosity of these channels and giving rise to a broader flow velocity distribution. At the molecular scale, the collision events between thiourea molecules present in the external mobile phase and the surface of the particles rarely lead to an effective absorption of thiourea into the particles. The molecules of thiourea that elute first (reflected by the sharp front of the black chromatogram in Figure~\ref{4}) are those mostly bouncing against the particle surfaces and transported along with fast velocity streamlines. Those eluting last (reflected by the rear tailing of the black peak in Figure~\ref{4}) are the few molecules localized within the particles that are slowly released and transported along with slow velocity streamlines.  

Even though the column performance is irreversibly affected by HPAM adsorption, the thiourea chromatograms measured in the HPAM-amended mobile phase at $\mathrm{Wi}>\mathrm{Wi}_\mathrm{c}$ (step \#3 in \S\ref{friction}) unambiguously demonstrate that the elastic instability promotes transverse dispersion in the column. The experimental results are summarized in Figure~\ref{5}; the normalized chromatograms obtained prior to and during flow of the HPAM-amended mobile phase are shown by the black and red curves, respectively. In the presence of the elastic instability, the elution peak shape returns to quasi-Gaussian (red curve), and the column efficiency more than doubles from $N_{\mathrm{Before}}$=160 to $N_{\mathrm{HPAM}}$=420---indicating that velocity biases are averaged by increasing the local transverse dispersion coefficient of thiourea ($D_t(x)$, Eq.~\ref{HeddyLong}, asymptotic dispersion regime). As a final confirmation of this point, we re-measure the thiourea chromatogram in polymer-free brine after washing and re-equilibrating the column with just brine (step \#4 in \S\ref{friction}). As shown by the blue curve in Figure~\ref{5}, the chromatogram returns to its broader peak shape with a lower efficiency $N_{\mathrm{After}}$=180, indicating that the increase in transverse dispersion is indeed generated by the elastic instability imparted by HPAM.

\section{Conclusion} 
The fundamental physics of elastic flow instabilities generated by flexible high molecular weight polymers has been studied for decades~\cite{larson_instabilities_1992,shaqfeh_purely_1996, pakdel_elastic_1996,groisman_elastic_2004,browne_bistability_2020,browne_elastic_2021,DattaReview,browne_homogenizing_2023,browne_harnessing_2023}. However, the use of such instabilities in applications has remained limited. Recent work~\cite{browne_harnessing_2023} showed that the chaotic flow generated by an elastic instability in model 3D glass bead packings---at $\mathrm{Re}\ll1$, where the flow is typically laminar---can dramatically enhance solute transverse dispersion, suggesting that such instabilities can help mitigate the negative impact of flow heterogeneities in chromatography. Here, we experimentally demonstrated a proof of this concept---for the first time, to the best of our knowledge. We generated the elastic instability by adding 500~ppm of 18~MDa HPAM polymer to a brine mobile phase in a 4.6~mm i.d. $\times$ 300~mm long HPLC column packed with 11.3~$\mu$m 125 \AA \ BEH fully porous organic/inorganic hybrid Particles. By measuring the friction factor of the column before, during, and after flow of the polymer-amended mobile phase, we determined the conditions under which the elastic instability arises. Then, by measuring the mass transfer of a non-retained small molecule analyte in the column before, during, and after flow of the polymer-amended mobile phase, we showed that the elastic instability causes the elution peak shape to become sharper and the column efficiency to more than double, owing to the increase of the transverse dispersion coefficient of the analyte across the column diameter. Our work thus paves the way towards closing the gap between the current efficiency of slurry-packed columns and the maximum theoretical efficiency of ideally-packed column without wall effects, simply by adding flexible polymers to the mobile phase. 

While our study highlights the promise of elastic instabilities in improving chromatography, this approach necessarily has limitations and challenges that remain to be addressed in future work to enable broad implementation. Most prominently, our measurements of the increased friction factor [red curve, Figure~\ref{3}(c)] and reduced column efficiency [Figures~\ref{4}--\ref{5}] indicate that the HPAM polymer irreversibly adsorbs to the external surfaces of the packed BEH125 Particles. Hence, more research is needed to develop ``polymer enhancers'' that can routinely be added to the mobile phase to minimize flow heterogeneities in chromatography with the following properties:
\begin{itemize}
    \item Sufficient elasticity to generate the flow instability, while still not generating a prohibitively large pressure drop across the column, thereby ensuring acceptable flow speeds in both preparative and analytical liquid chromatography in addition to having a practically-feasible onset flow rate to generate the elastic instability.
    \item Compatibility with the use of common mobile phase compositions in RPLC, HILIC, IEX, and SEC. In this work, for instance, no organic modifier was added to the mobile phase while it is commonly used in RPLC and HILIC.
    \item Minimal irreversible adsorption to the surfaces of the packed particles, and similarly, chemically inert with respect to the analyte.
    \item Minimal scission of the polymer chains as they are transported through the tortuous pore space.
    \item Does not interfere with analyte detection (via optical, mass, and fluorescence means). Here, the HPAM polymer was suitable for optical detection wavelengths only larger than 270 nm. 
    \item Increases transverse dispersion via an elastic instability, but does not simultaneously increase axial dispersion to a prohibitive extent.
\end{itemize}

\section{Acknowledgements} 
It is a pleasure to acknowledge funding support from the Princeton Center for Complex Materials (PCCM), a National Science Foundation (NSF) Materials Research Science and Engineering Center (MRSEC; DMR-2011750), as well as the Camille Dreyfus Teacher-Scholar Program. We also acknowledge the use of Princeton’s Imaging and Analysis Center (IAC), which is partially supported by the PCCM.

\section{Note}
BEH, ACQUITY, UPLC, Arc, and Empower are trademarks of Waters Technologies Corporation.

\newpage

\newpage

\newpage

\newpage

\newpage

\end{document}